\documentstyle[sprocl,epsfig]{article}

\def\be{\begin{equation}}
\def\ee{\end{equation}}
\def\bea{\begin{eqnarray}}
\def\eea{\end{eqnarray}}

\begin{document}

\title{GAMMA RAY BURSTS: SOME FACTS AND IDEAS}

\author{GABRIELE GHISELLINI}

\address{Osservatorio Astronomico di Brera, Via Bianchi 46, I--23807 Merate
Italy}


\maketitle
\abstracts{Gamma Ray Bursts (GRBs) are the most explosive
events after the big bang: their energy output corresponds 
to a sizeable fraction of a solar mass entirely converted 
into energy in a few seconds.
Although many questions about their progenitors remain
to be answered, it is likely that they are generated 
by a newly formed and fast spinning black hole.
The colossal power characterizing GRBs
is carried by a surprisingly small amount of matter,
which is accelerated to speeds differing from $c$ by one part 
in ten thousands.
GRBs are then the most (special and general) relativistic
objects we know of. 
Since GRBs are the brightest sources at high redshift, albeit
for a limited amount of time, they are also the best torchlights
we have to shine the far universe.
}

\section{Introduction}
On July 2, 1967, the american VELA{\footnote{Note that the name 
VELA stands for ``to watch", from the spanish {\it velar.}}}
satellites detected a flash of mysterious gamma rays 
coming from space.
Soon after, other events, similar in nature and duration,
were detected, even if the announcement of this major
discovery awaited a few years, when was finally reported
by Klebesadel, Strong \& Olson (1973).
Gamma--ray bursts (GRB) were therefore discovered by accident,
thanks to a series of small military satellites designed to
detect the radiation produced by the explosion
of thermo--nuclear bombs, that had recently being banned
by an international treaty.
Even if small X--ray detectors could be enough for the detection of nuclear 
bomb exploding on earth, it was thought that the russians could have 
the technology to let a bomb explode on the dark side of the moon.
In this case the X--rays are completely shielded, but the 
radio--active blast triggered by the bomb and expanding beyond the lunar
dish, produces detectable $\gamma$--rays.
This was the reason for having small $\gamma$--ray detectors on board.
The VELA (limited) capability to reconstruct the source direction 
through differences in the arrival photon time (there always were two
or more satellites in orbit) enabled to exclude the earth, the moon and
the sun as sources, and established the cosmic origin of GRBs
(see J. Bonnell at 
{\tt http://antwrp.gsfc.nasa.gov/htmltest/jbonnell/www/grbhist.html}).

30 years since their detection by the VELA satellites,
we now start to understand the physics of GRBs.
This has been made possible by the precise localization of the
Wide Field Camera of $Beppo$SAX, which allowed the detection of their
X--ray afterglow emission (Costa et al. 1997) and the optical follow up
observations, leading to the discovery that they are cosmological sources 
(van Paradijs et al. 1997).
The huge energy and power releases required by their cosmological
distances support the fireball scenario (Cavallo \& Rees 1978; Rees
\& M\'esz\'aros 1992; M\'esz\'aros \& Rees 1993),
even if we do not know yet which kind of progenitor makes the GRB phenomenon.

There are already excellent reviews on GRBs (van Paradijs, 
Kouveliotou, \& Wijers 2000; Meszaros 2001, Pian 2001,
Klose 2000, Piran 1999, Meszaros 1999, Fishman \& Meegan 1995),
and I will not even try to be exhaustive in this contribution.
I will instead concentrate on non--technical aspects of  
this very active research field, aiming at non specialized readers,
and to point out some of the problems that are
still under discussion.

\begin{figure}
\vskip -0.5 true cm
\centerline{\psfig{figure=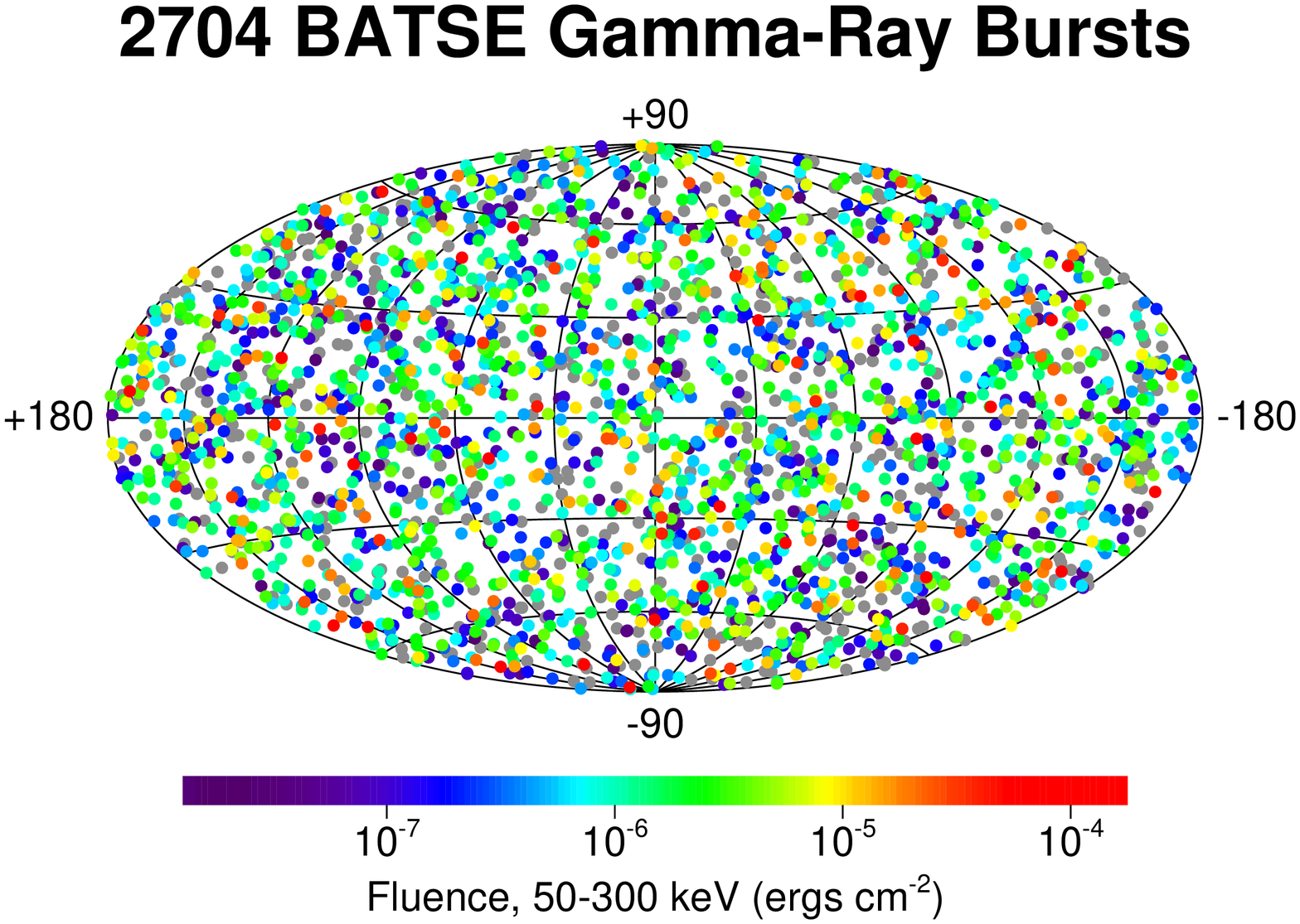, width=9cm, angle=90}} 
\vskip -0.5 true cm
\caption{Distribution in the sky of the GRBs seen by BATSE.
The projection is in Galactic coordinates.
Grey levels corresponds to different fluences, i.e. the integral 
of the burst flux over time.
Note the very high level of isotropy.
(from {\tt http://www.batse.msfc.nasa.gov/batse/grb/skymap/})
}
\end{figure}

\section{Why GRBs are so fascinating?}

Since their discovery, GBRs acquired a special ``aura", due, I think, 
mainly to their mysterious nature which persisted for decades.
This certainly attracted many astrophysicists looking for a well 
posed but unsolved problem, and part of the popularity
of GRBs is still due to that.
The mystery lasted for so long because it was not possible to
detect any counterpart at any wavelength, due to the too large
error boxes of the positions given by the $\gamma$--ray instruments.
Now, after knowning about their cosmological nature, the
importance of GRBs has, if possible, increased, because they
are good probes in at least three broad areas of physics:
general relativity, special relativity, and cosmology.
In fact we suspect they flag the birth of 
a stellar mass black hole, and that the corresponding liberation
of energy is able to accelerate matter to bulk Lorentz factors
exceeding 100, equivalent to speeds differing from $c$ by less than
one part in ten thousand.
When shining, they are the most powerful objects of the 
universe, and even if their brightness decreases fast, they 
offer the opportunity to explore the far universe  (i.e. redshift 
greater than 10, if such distant bursts exist) with a level of 
detail impossible to achieve with any other class of objects.

\begin{figure}
\vskip -0.7 true cm
\centerline{\psfig{figure=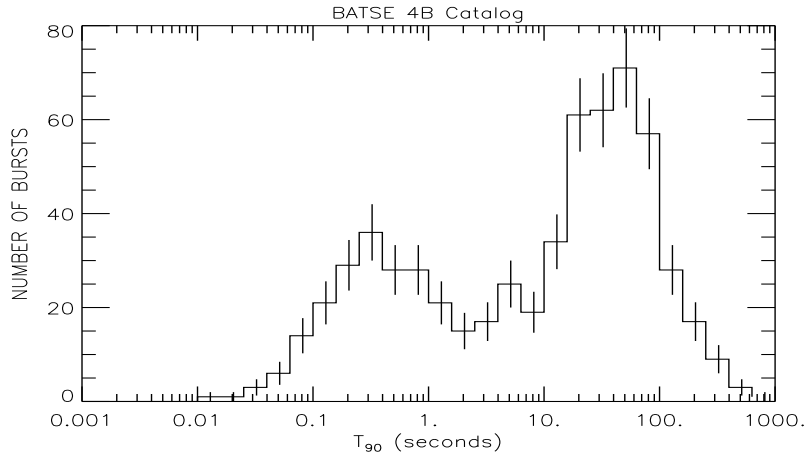, width=11cm, height=6cm}}
\vskip -0.3 true cm
\caption{Distribution of durations of GRBs.
Shown here is the ``$T_{90}$" duration, meaning the 
interval of time over which a burst emits from 5\% of 
its total measured counts to 95\%. 
The counts are integrated over all the 4 BATSE channels (i.e. $E > 20$ keV). 
(From {\tt http://www.batse.msfc.nasa.gov/batse/grb/duration/)}
}
\end{figure}

\section{Facts I: the pre--BeppoSAX era}
After the discovery phase and after the confirmation by many
small satellites hosting on board $\gamma$--ray
detectors, we entered in the {\it Compton Gamma Ray Observatory (CGRO)}
satellite era.
Launched in April 1991, it had onboard several high energy 
instruments, and in particular BATSE, made by 8 $\gamma$--ray 
detectors (sensitive in the 20--600 keV range) especially 
designed for GRB observations, and EGRET, sensitive above 100 MeV.

\vskip 0.3 true cm
\noindent
{\bf Isotropy in the sky ---} 
Fig. 1 shows the positions of 2704 bursts detected by BATSE,
and it can be seen that their locations are distributed completely
isotropically in the sky. No dipole or quadrupole moments have been
detected.

\vskip 0.3 true cm
\noindent
{\bf Duration ---} 
Fig. 2 shows that GRBs seem to come in (at least) two flavors:
the majority of them lasts for more than 2 seconds, while
about one third is shorter.
All information derived from the precise localization of GRBs refer to 
{\it long} bursts.
The bimodality of the distribution of their duration (measured by
the time $T_{90}$ within which 90 per cent of the total fluence is
contained) is confirmed by the associated spectral shape,
since short bursts, on average, appear harder than long GRBs. 
Fig. 5 shows the hardness ratio (which is a measure of the 
slope of the spectrum: larger values means that the flux
at high energies is more dominating) as a function of the duration
of the emission.
\begin{figure}[h]
\begin{tabular}{cc}
\psfig{figure=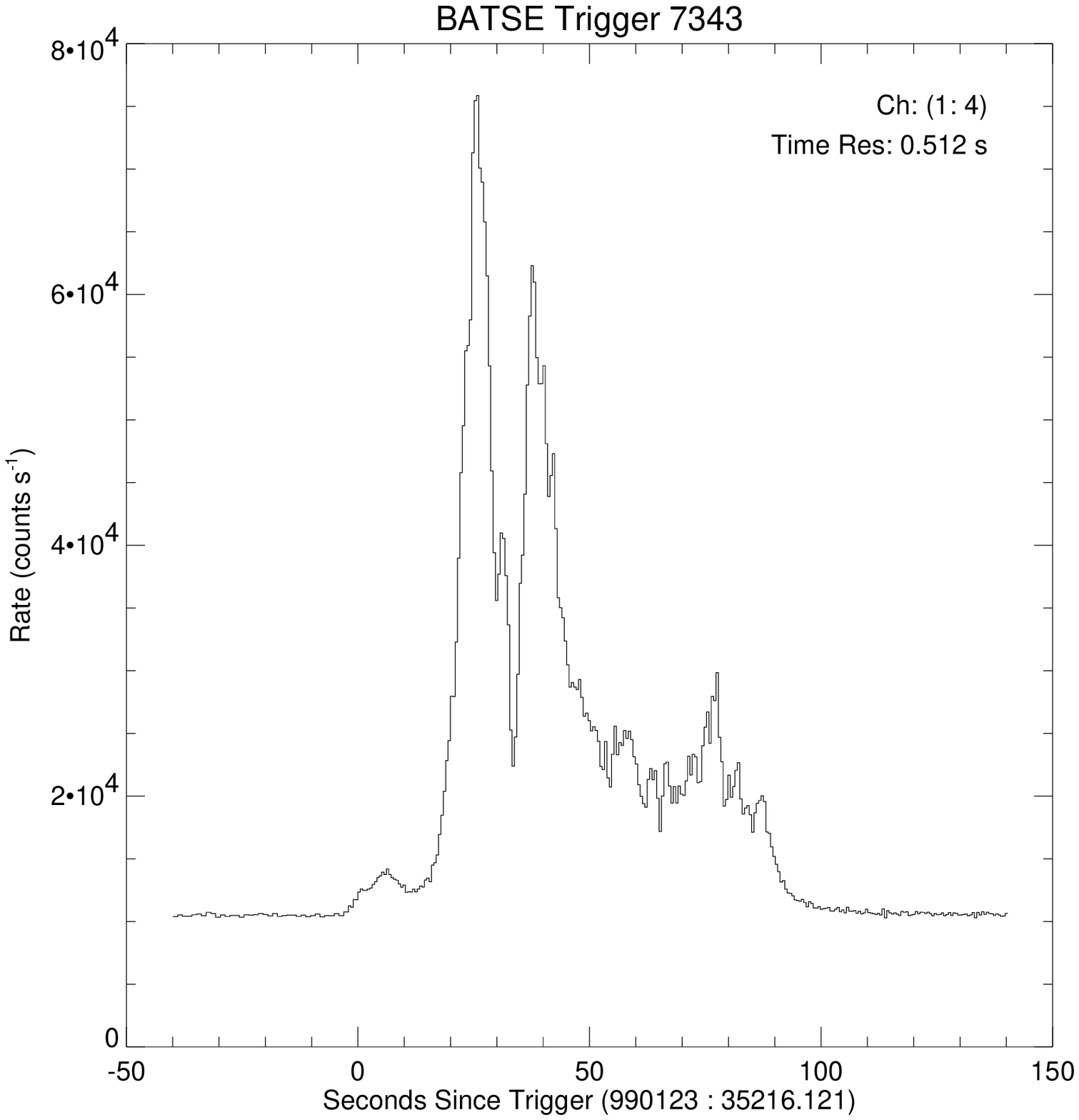, width=5.5cm, height=4.7cm} 
&\psfig{figure=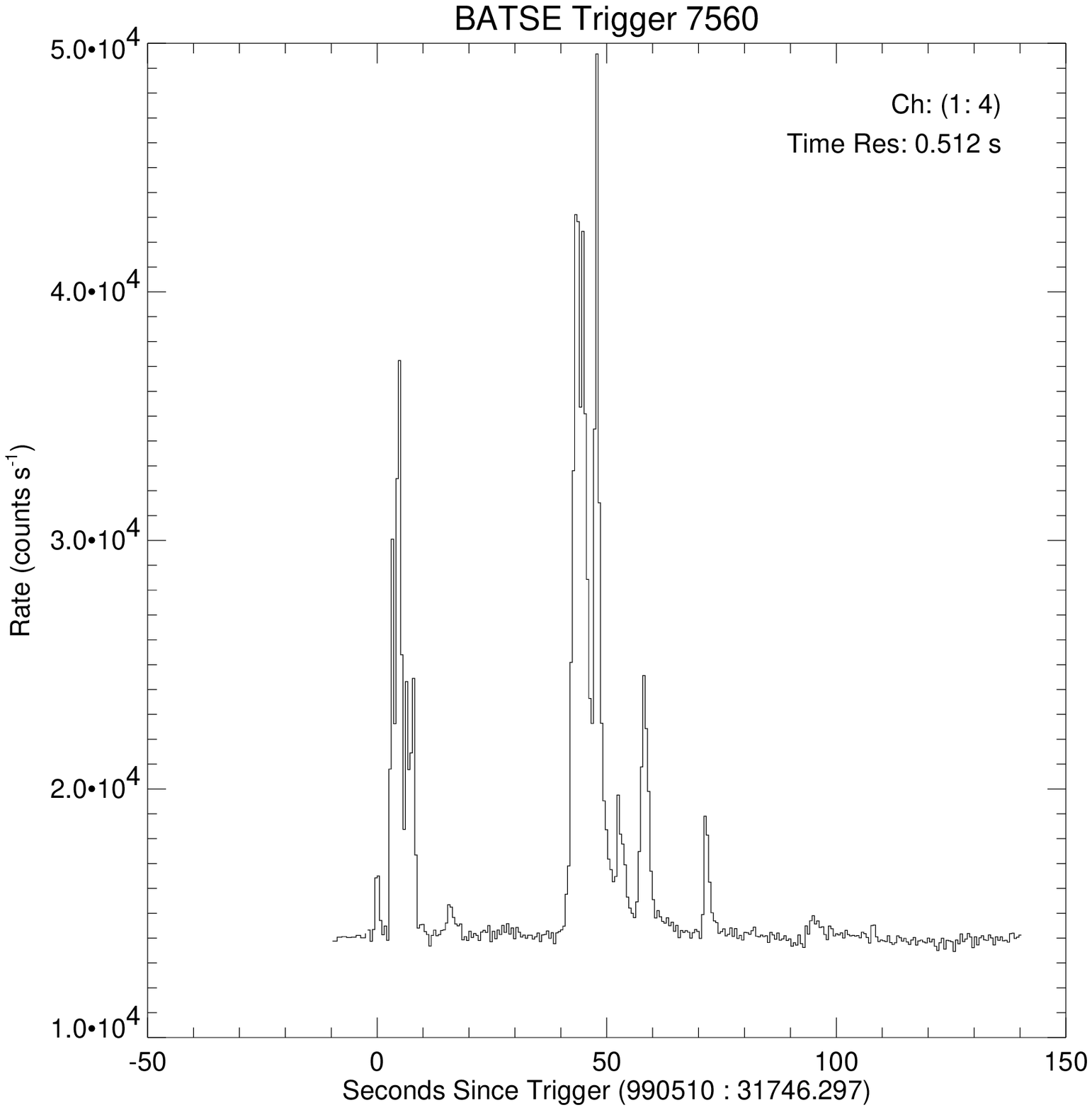, width=5.5cm, height=4.7cm} \\
\psfig{figure=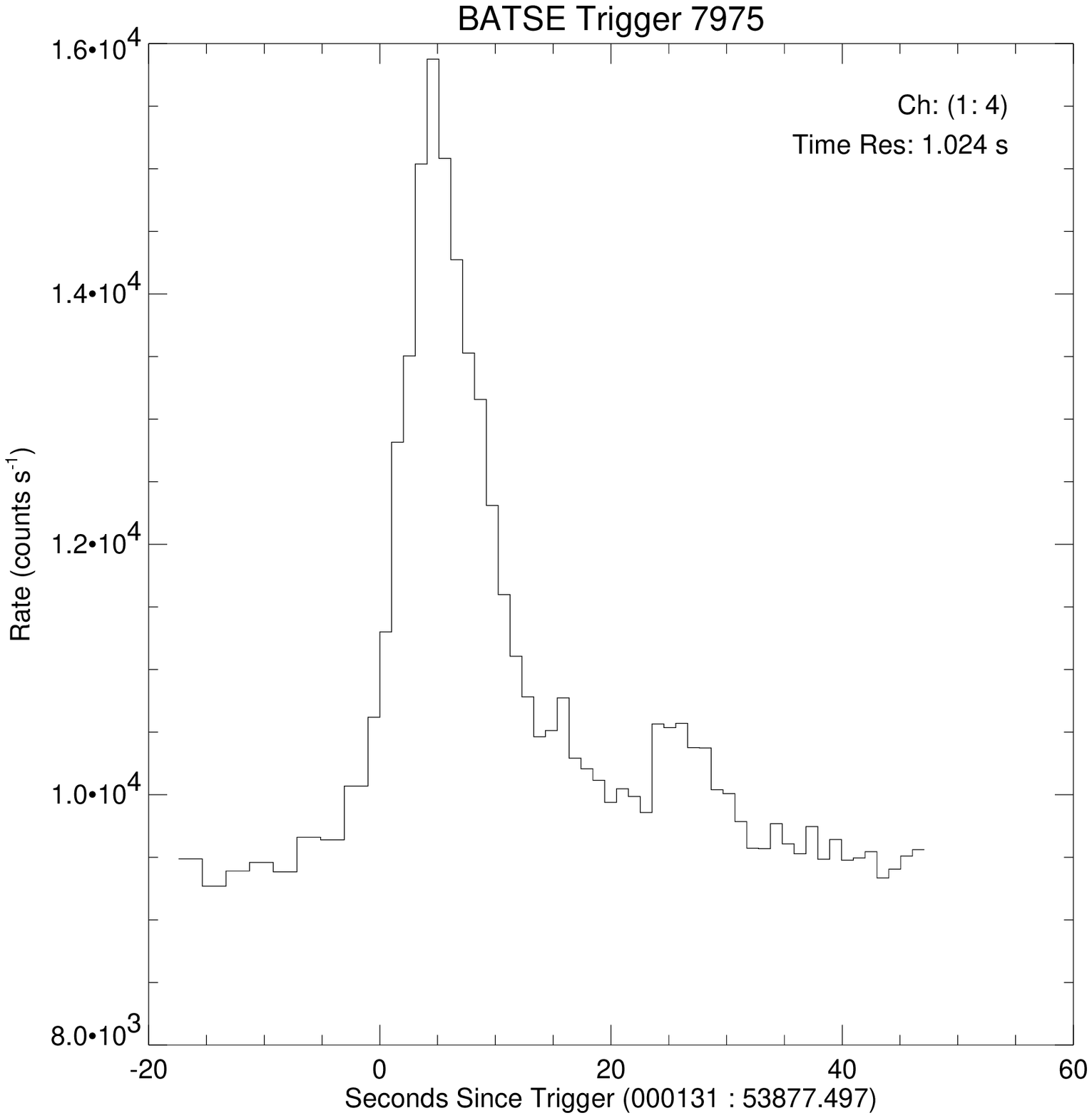, width=5.5cm, height=4.7cm}
&\psfig{figure=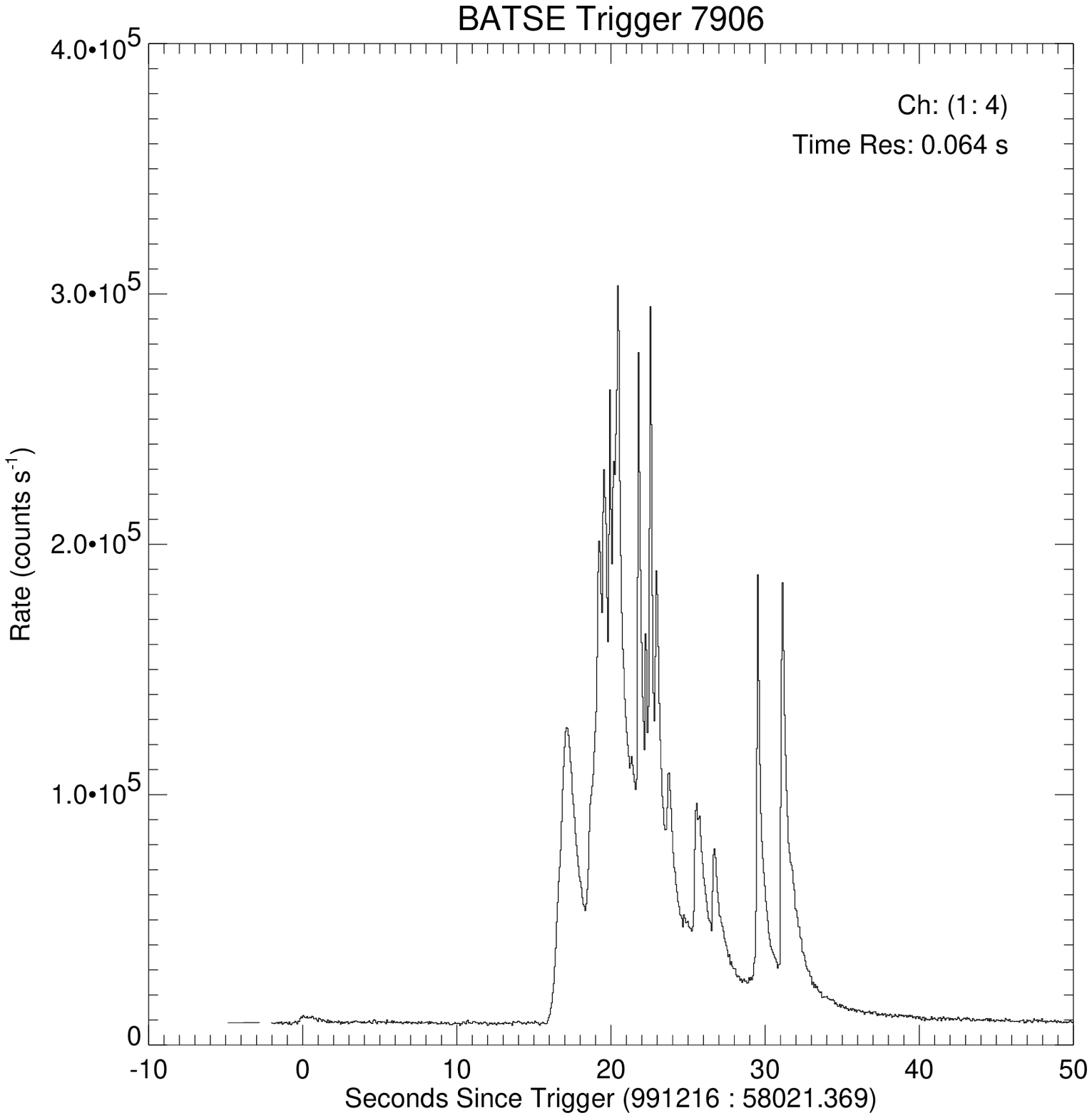, width=5.5cm, height=4.7cm} 
\end{tabular}
\caption{BATSE light curves of some GRBs.
Clockwise, from top left: GRB 990123, GRB 990510, GRB 991216 and GRB 000131.}
\end{figure} 
\begin{figure}
\vskip -3.8 true cm
\centerline{\hskip 2.5 true cm \psfig{figure=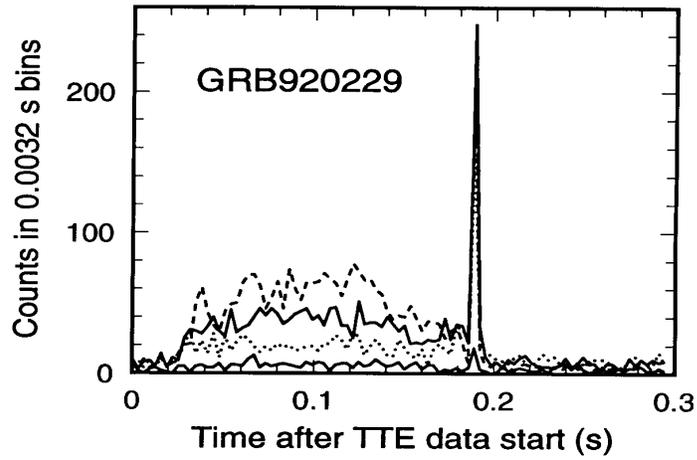, width=13cm, height=10cm}}
\vskip -0.5 true cm
\caption{BATSE light curve of GRB 920229, showing the spike
of 0.22 millisecond. 
Different curves corresponds to the different BATSE energy channels.
From Schaefer \& Walker (1999).}
\end{figure}
\begin{figure}
\centerline{\psfig{figure=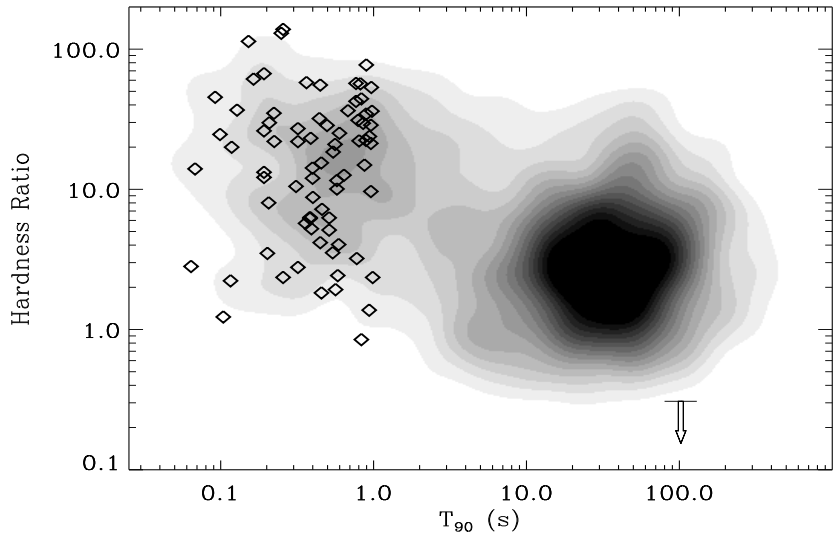, width=11cm, height=7cm}}
\caption{Hardness ratio vs duration of BATSE bursts.
The hardness ratio is a measure of the shape of the spectrum:
larger values correspond to harder spectra.
Different grey levels are density contours.
The diamonds mark the bursts used for the search of the X--ray afterglow
in short burst made by Lazzati, Ramirez--Ruiz \& Ghisellini (2001)}
\end{figure}

The light curve of GRBs is erratic (see Fig. 3) and sometimes highly
variable: spikes as short as a fraction of a millisecond have 
been detected (see Shaefer \& Walker 1999 and Fig. 4).
The issue of variability is central for the modeling of GRBs:
the extremely short timescales we observe demand 
large Lorentz factors, and the fact that the spikes
at early and late times of the prompt emission have
similar timescales (i.e. their duration does not increase) are
major proofs against external shocks (see below) causing
the prompt emission of GRBs (Fenimore, Ramirez--Ruiz \& Wu 1999).

In addition, as Amelino--Camelia et al. (1998) pointed out (see also these 
proceedings) the very short variability of high energy photons
coming from a cosmological (i.e. redshift greater than one)
source can carry key information about the structure of spacetime, 
which can limit the possibility of having a breaking of 
the Lorentz invariance as proposed by some theories.

\vskip 0.3 true cm
\noindent
{\bf Fluences ---} 
Most $\gamma$--ray fluences (i.e. the flux integrated over the duration
of the burst) are in the range $10^{-6}$--$10^{-4}$ erg cm$^{-2}$.
The number counts of bursts are very flat, 
with $\langle V/V_{max}\rangle \approx 1/3$.
Schmidt (2001) finds that short and long bursts have similar values
of $\langle V/V_{max}\rangle$, while Tavani (1998) found that short 
and {\it soft} bursts show little deviation from the Euclidean value 
$\langle V/V_{max}\rangle=0.5$.

\vskip 0.3 true cm
\noindent
{\bf Spectra ---} 
The spectra of GRBs are very hard, with a 
peak (in a $E$--$E F_E$ plot) at 
an energy $E_{peak}$ of a few hundreds keV.
Fig. 6 shows the distributions of the photon spectral indices
derived by fitting the spectrum with the Band function
(Band et al. 1993), consisting in two smoothly connected 
power laws (from Lloyd \& Petrosian 1999, 2000, see also 
Preece et al. 1998), defined as $N(E) \propto E^{\alpha}$ at
low energies and $N(E) \propto E^{\beta}$ at high energies.
As can be seen, the $\alpha$--distribution peaks in the range
[$-$1, $-$0.5] corresponding to $F(E) \propto E^0$--$E^{1/2}$.
This spectral index varies during the burst, as does $E_{\rm peak}$.
The general trend is that the spectrum softens, and $E_{\rm peak}$
decreases, with time. 
More precise statements must however wait for larger area detectors,
since what we inevitably do, at present, is to fit a time integrated 
spectrum of a very rapidly variable source:
the minimum integration time is $\sim$0.1 second for the strongest bursts,
while the variability timescales can be hundreds of times shorter.
Some bursts have been detected at very large $\gamma$--rays energies 
($>$ 100 MeV) by the EGRET instrument (see the review by Fishman 
\& Meegan 1995, and references therein).

\begin{figure}
\vskip -1 true cm
\centerline{\psfig{figure=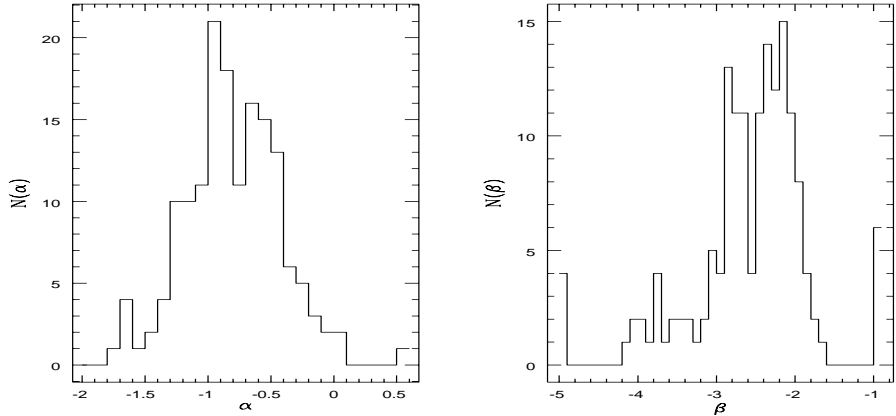, width=6cm, height=13cm, angle=-90}}
\caption{The distributions of the (photon) energy indices
$\alpha$ and $\beta$ characterizing the Band function (Band et al. 1993).
$N(E)\propto E^{\alpha}$ at low energies and $\propto E^{\beta}$
at high energies. From Lloyd \& Petrosian 2000.}
\end{figure}

\section{Facts II: the BeppoSAX era}

Launched on April 1996, the {\it Beppo}SAX satellite made the
breakthrough observations of GRBs, succeeding in positioning 
them with error boxes of only a few arcminutes through its
coded mask Wide Field Camera sensitive at medium--hard X--ray energies
[2--25 keV].
This made it possible to promptly slew the satellite in the
found direction, to observe the error box region with the
Narrow Field Instruments (NFI, in the 0.1--10 keV band) 
and to detect (in all cases but one) a new fading X--ray source ,
immediately identified as the X--ray afterglow of the GRB.
Thanks to the prompt dissemination of the coordinates through the 
Gamma--ray Burst Coordinates Network (i.e. the GCN system,
created by Scott Barthelmy) all ground based telescopes can point 
the target and try to detect the optical (and IR, and radio) afterglow.

\vskip 0.3 true cm
\noindent
{\bf X--ray afterglows ---}
Usually, {\it Beppo}SAX could re--point in 6--8 hours from the 
trigger, and detect the X--ray afterglow,
at an initial level of $\sim 10^{-13}$ erg s$^{-1}$ cm$^{-2}$ or higher, 
and with a flux decaying roughly as $F_X(t) \propto t^{-1}$ -- $t^{-1.5}$.
The spectral shape is remarkably softer than the prompt emission,
with $F(E)$ roughly proportional to $E^{-1}$.
When does the afterglow begins?
We do not know yet, and this is one of the key question that
HETE II and especially Swift can answer.
Some hints (from SIGMA: GRB 920723, Burenin et al. 1999; 
from BATSE: GRB 980923, Giblin et al. 1999; 
from {\it Beppo}SAX, Frontera et al. 2000) 
suggest that the afterglow starts during the prompt emission or immediately 
after (and this would be in agreement with the optical flash observed 
in GRB 990123), but much more firm evidence is needed.

\vskip 0.3 true cm
\noindent
{\bf Optical afterglows ---} 
For about 1/2 of the bursts
with good locations an optical afterglow has been detected. The
monochromatic flux decreases in time as a power law 
$F_\nu(t)\propto t^{-0.8}$--$t^{-2}$.
Usually, the magnitudes of the optical afterglow detected
$\sim$one day after the $\gamma$--ray event are in the range 19--21. 
GRB 990123 is still the only burst detected so far in the optical
while the prompt emission was still on,  by
the robotic telescope ROTSE, 22 seconds after the $\gamma$--ray
trigger at $m\sim 11.7$, reaching $m\sim 8.9$ 47 seconds after the
trigger (Akerlof \& McKay 1999; Akerlof et al. 2000) 
{\footnote{This is the {\it optical flash}, and is interpreted by the
emission of the reverse shock crossing the fireball when first
impacting into the interstellar material.
The term {\it flash} is indeed appropriate.
A magnitude 9 at a redshift $z=1.6$ corresponds a power $L\sim 5\times 10^{49}$
erg s$^{-1}$ in the optical band, meaning that if the same kind of object switches 
on at 1 kpc from earth, we would have {\it two equally bright Suns} in the sky, 
albeit for a short time.}}.
For a complete review about the issue of optical and infrared
afterglows see Pian (2001).

More than half of the bursts whose X--ray afterglow is detected 
could not be detected at optical frequencies, despite many
of them have been observed with large telescopes at early times
(see below and Lazzati, Covino \& Ghisellini 2002).

\begin{figure}
\vskip -1 true cm
\psfig{figure=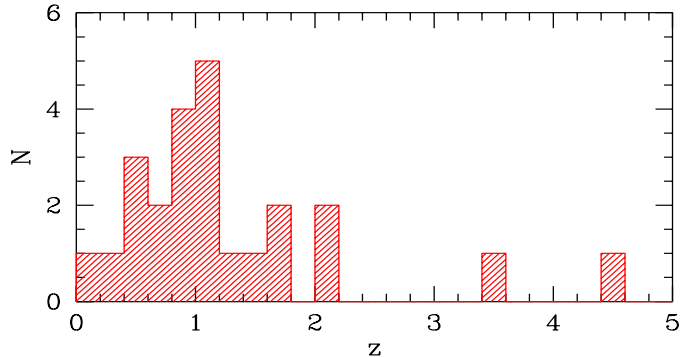, width=10cm, height=9cm}
\vskip -3.5 true cm
\caption{Distribution of all the known redshifts of GRBs as of
November 2001. All redshifts are within the 0.4--4.5 range,
apart from GRB 980425: if it is really associated with SN1998bw,
then its redshift is $z=0.008$.}
\end{figure}
\begin{figure}
\vskip -1 true cm
\centerline{\psfig{figure=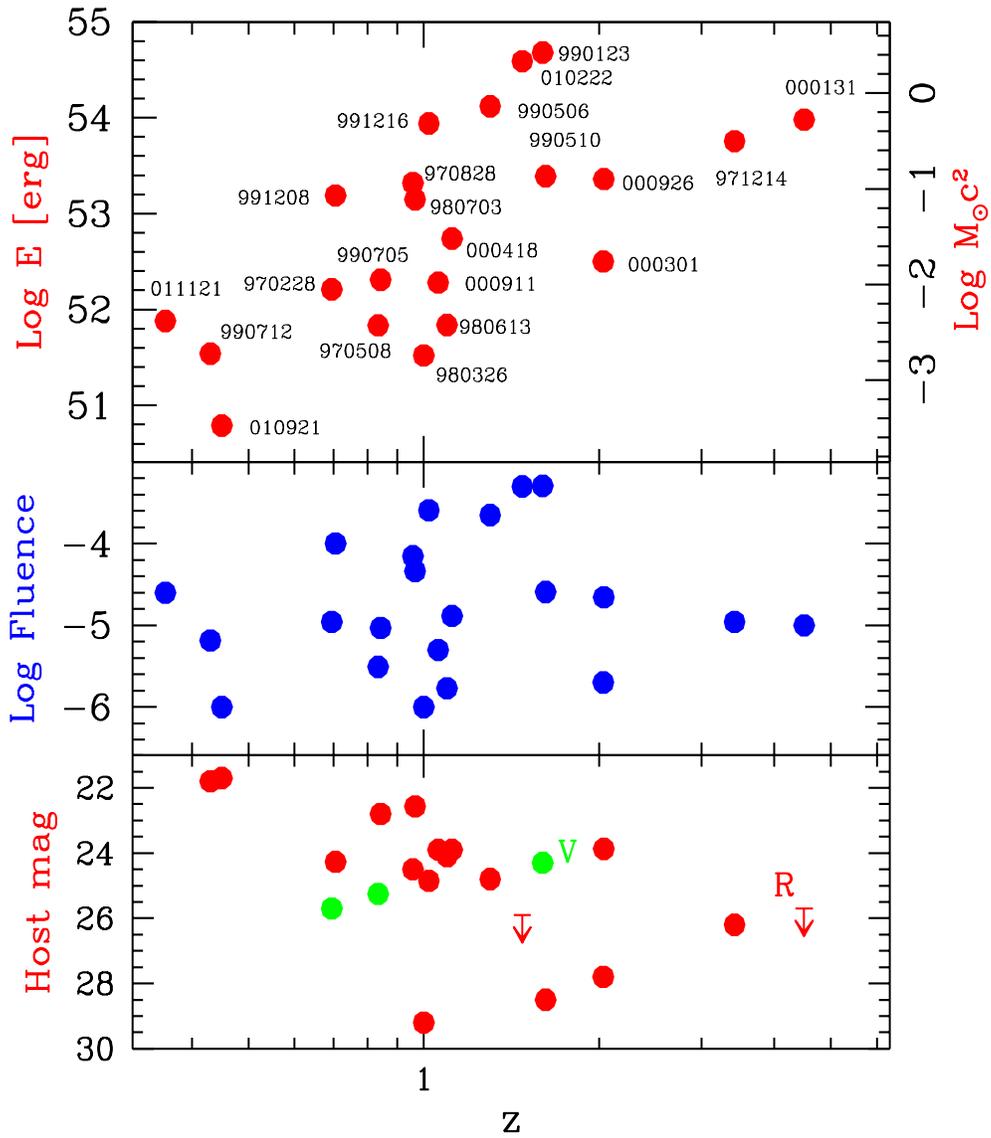, width=15cm, height=16cm}} 
\vskip -0.7 true cm
\caption{Energy, fluence and host magnitude as a function of redshift
for all bursts of known redshift, as of November 2001.
The energy is calculated assuming isotropic emission.
Most of the magnitudes of the host galaxy are in the $R$ filter,
a few in the $V$ filter (as labeled).
}
\end{figure}

\vskip 0.3 true cm
\noindent
{\bf Radio afterglows ---} 
The first GRB radio afterglow was detected for GRB 970508.
In this case, the radio flux at 8.5 GHz  behaved ``nervously" 
for about one month, to ``calm down" after this time.
This has been interpreted as due to interstellar scintillation
in our Galaxy, affecting sources of very small angular diameter.
Therefore the fact that scintillation ceased after some
time was a sign of the increased source dimension.
Since this is a burst of known distance, it is possible to estimate 
the size needed to quench the scintillation effect, 
and then to estimate the velocity of the fireball, which
turned out to be relativistic (Frail et al. 1997).

\vskip 0.3 true cm
\noindent
{\bf Redshifts ---}
Up to November 2001, about 20 redshifts of GRBs have been measured,
and Fig. 7 shows their distribution.
Apart from the controversial case of GRB 980425, possibly
associated with the nearby SN 1998bw (at $z\sim 0.008$),
all other redshifts are within the 0.4--4.5 range.
Fig. 8 shows the $\gamma$--ray luminosity (assumed isotropically
emitted), the  $\gamma$--ray fluence and the magnitude of the 
host galaxy (when detected) as a function of redshift.
A particularly useful updated link with all the relevant information
about bursts with good localization is maintained by Jochen Greiner at:\\
{\tt http://www.aip.de/{$\sim$}jcg/grbgen.html}

\begin{figure}
\vskip -0.5 true cm
\centerline{\psfig{figure=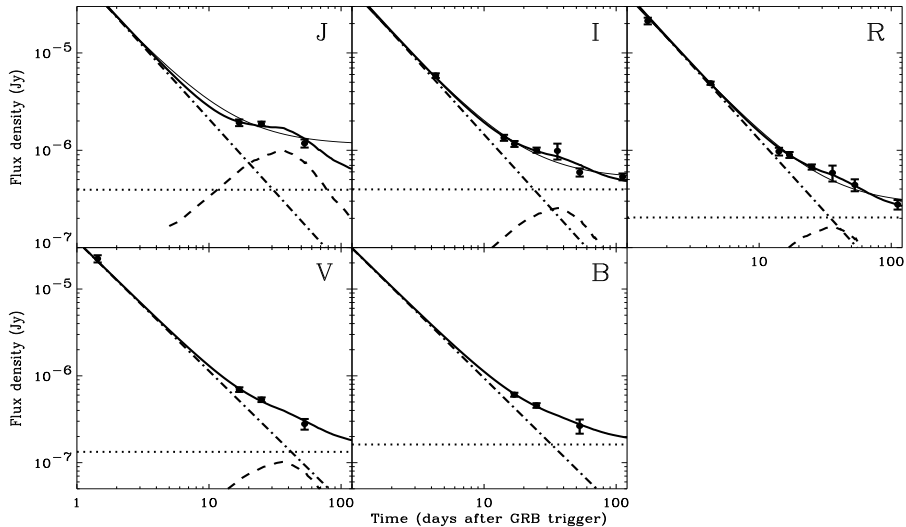,width=12cm, height=7cm}}
\caption{The late rebrightening of the IR--optical light curve can flag 
the presence of an underlying supernova, as suggested for GRB 000911 by 
Lazzati et al. (2001).
The horizontal line is the (constant) contribution of the host galaxy,
the bumpy curve is the assumed supernova contribution,
more visible in the infrared ($J$ filter).}
\end{figure}

\vskip 0.3 true cm
\noindent
{\bf The host galaxies ---}
The host galaxies of GRBs are not particularly luminous ($L<L_*$),
they appear blue (Hogg \& Fruchter 1999), and the location of the optical 
transient is never very distant from the galaxy center, in agreement
with the idea that long GRBs are associated with massive (and
short lived) progenitors (for a recent review see Djorgovski et al. 2001).
Fig. 8 shows the magnitude of the host as a function of redshift
(almost all magnitudes are in the $R$ band).

\vskip 0.3 true cm
\noindent
{\bf The GRB--Supernova connection ---}
It is possible that GRB 980425 was associated with the
nearby supernova SN 1998bw.
If so, this GRB would be very anomalous, being 3--4 orders of 
magnitude underluminous with respect to the other bursts of known redshift.
On the other hand even SN 1998bw is far from being a typical 
member of its class (for instance, from its radio light curve 
Kulkarni et al. 1998 estimated a very high brightness temperature,
requiring a bulk flow with $\Gamma\sim 2$).
Independent of the reality of this association, these observations
triggered much interest about the possibility that GRB are 
in some way associated with some rare kind of supernovae.
The other main evidence in favor of it is the re--brightnening of the late
optical--infrared light curve observed in some afterglows
(Bloom et al. 1999; Reichart 1999; Galama et al. 2000; 
Lazzati et al., 2001), especially if accompanied by evidence of a
spectral change (spectrum which becomes redder), as appropriate
if the supernova emission exceeds the afterglow light
(see one example in Fig. 9).

\vskip 0.3 true cm
\noindent
{\bf Iron lines ---}
There are five bursts displaying evidence
for large amounts of X--ray line emitting material around the site 
of the explosion: four 
(GRB 970508, Piro et al., 1999;
GRB 970828, Yoshida et al., 1999; 
GRB 991216, Piro et al., 2000; 
GRB 000214, Antonelli et al., 2000) 
show an emission feature during the afterglow, and one 
(GRB 990705, Amati et al., 2000) 
displays an edge in absorption during the burst itself. 
The detection of emission features in the afterglow spectra of 
GRBs some hours after the GRB event poses strong constraints
on the properties of the line--emitting material.

The first constraint comes from the large detected flux. 
Assuming the line lasts for $10^5 t_5$ seconds at the
level of $10^{-13}F_{Fe,-13}$ erg cm$^{-2}$ s$^{-1}$ for a burst
at $z=1$, this corresponds to $\sim 3\times 10^{57}$ line photons.
If each iron atom produces $k$ line photons, the required total mass
of iron is $M_{Fe}\sim 150 F_{Fe,-13}t_5/k$ solar masses.
It is clear that we need at the very least $k>10^3$ to bring
$M_{Fe}$ down to reasonable values.
This means that the iron must recombine sufficiently fast, 
and this implies large densities and not so large temperatures.

The second constraint concerns the geometrical setup: the line
emitting material cannot be between us and the burst, because 
1) we should see an absorption line, and 2) the large amount 
of material would stop the fireball and the associated afterglow,
contrary to what observed.
The line emitting material must therefore be out of the line of sight
and nevertheless be illuminated by a large amount of ionizing flux.
Furthermore, for a line observed after a time $\Delta t$ since trigger, 
we must have that the emitting material must be close to the burst site,
within a distance $R$ given by
\begin{equation}
R\, \le\, {c\Delta t \over 1+z} \, {1\over 1-\cos\theta}\, \simeq \,
{1.1\times 10^{15}\over 1+z}  \, {\Delta t \over 10\, {\rm h} } \,\, 
{1\over 1-\cos\theta}\,\, {\rm cm},
\end{equation}
where $\theta$ is the angle between the location of the material 
and the line of sight (see the cartoon in Fig. 15 showing some 
possible set--ups, and see below for a brief discussion of
the proposed models).

\begin{figure}
\vskip -0.5 true cm
\centerline{\psfig{figure=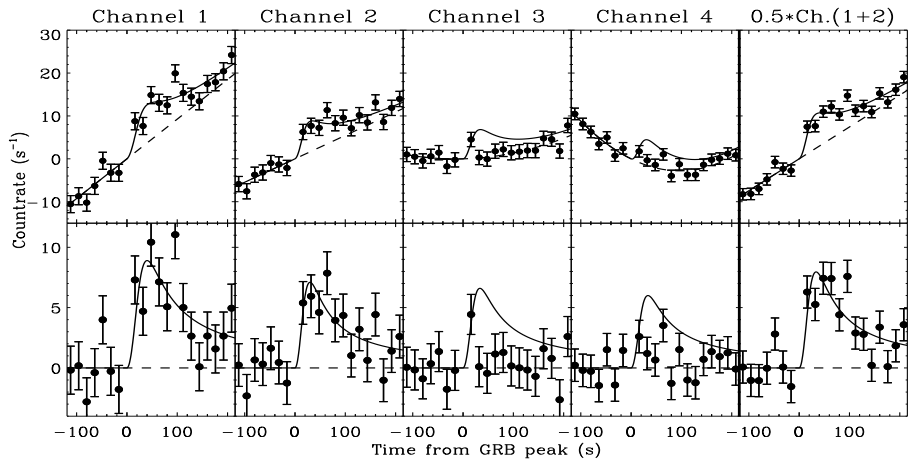, width=12cm, height=6cm}}
\caption{X--ray afterglow for the short bursts? The upper panels
show the sum of the BATSE light curves of the selected short
bursts (in each channel, while the rightmost
panel shows the sum of the first and second channels).
The time bin corresponding to the burst emission has been removed.
The bottom panels show the background subtracted light curves.
Note the excess at 30--100 seconds in the first two channels.
The significance of this excess is $\sim3.5\sigma$ for the signal 
in individual channels and $\sim 4.2\sigma$ for the sum of the 
first and second. From Lazzati, Ramirez--Ruiz \& Ghisellini (2001).}
\end{figure}

\subsection{Afterglow in short GRBs?}

Lazzati, Ramirez--Ruiz \& Ghisellini (2001) have summed up
the BATSE light curves of the best signal--to--noise bursts lasting 
for less than one second. 
The resulting 76 light curves were analized separately in each
of the four BATSE energy channels, and some excess was found in the 
first two BATSE channels, between 30 and 100 seconds after the trigger.
The four flux data points (two detections and two upper limits)
indicate a relatively $steep$ spectrum, similar to the shape
of the X--ray afterglow of long bursts.
This could be the first detection of an afterglow in
short bursts: if confirmed, it proves that also short bursts
can transform bulk into random energy, suggesting the presence
of a relativistic fireball also in this case.

\section{The fireball}

The energy involved in GRB explosions is huge.
No matter in which form the energy is initially injected,
a quasi--thermal equilibrium (at relativistic temperatures)
between matter and radiation is reached,
with the formation of electron--positron pairs accelerated to
relativistic speeds by the high internal pressure.
This is a {\it fireball} (Cavallo \& Rees 1978).
When the temperature of the radiation (as measured in the comoving
frame) drops below $\sim$50 keV the pairs annihilate 
faster than the rate at which they are produced.
But the presence of even a small amount of barions, corresponding to 
only $\sim 10^{-6}~ M_\odot$,
makes the fireball opaque to Thomson scattering: the internal radiation
thus continues to accelerate the fireball until most of its initial 
energy has been converted into bulk motion. 
After this phase the fireball expands at a constant speed and at some point 
becomes transparent.

\begin{figure}
\vskip -0.5 true cm
\centerline{\psfig{figure=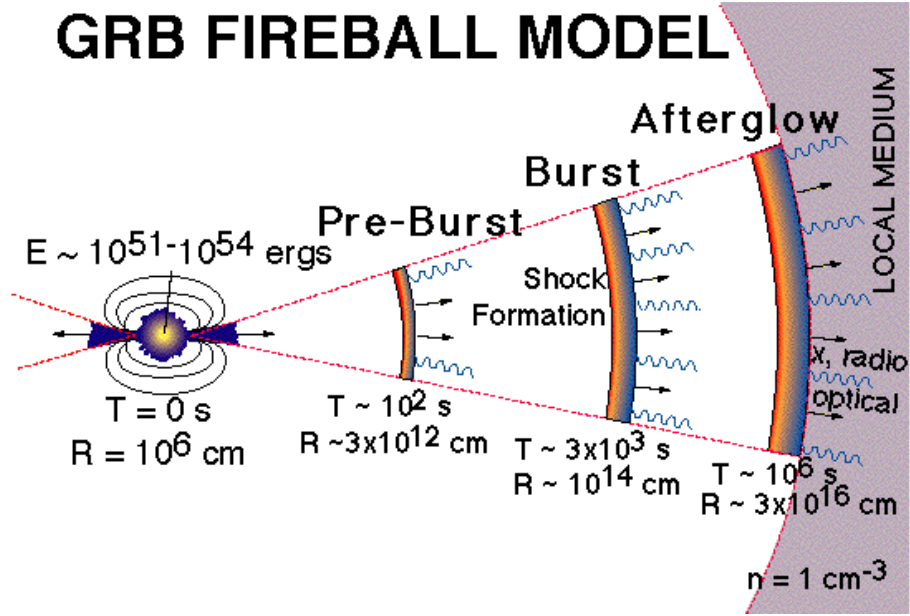, width=12cm}} 
\caption{Cartoon for the fireball and the internal/external shock scenario.}
\end{figure}

\vskip 0.3 true cm
\noindent
{\bf The compactness problem ---} Even before the discovery that 
GRBs are at high redhift, researchers were puzzled about 
such rapidly variable and strong high energy fluxes.
In fact, even if bursts were close by (say, in the Galactic halo),
they would be highly super--Eddington, and this poses the problem
to explain how high energy $\gamma$--rays can survive against
the $\gamma$--$\gamma\to e^{\pm}$ process. 
In fact, from the minimum variability time--scale
(time needed to double or halve the flux) we can estimate, 
by a causality argument, the size $R$ of the emitting region.
Therefore we can form the luminosity to size ratio $L/R$ which 
controls the processes involving photons, and in particular
the $\gamma\gamma\to e^\pm$ process
(in fact its optical depth $\tau_{\gamma\gamma}\propto R (L/R^2)\propto L/R$.
Taken at face value, the $L/R$ ratio is too large 
(and therefore GRBs are {\it too compact}) 
to let any photon above threshold for pair production 
(i.e. with an energy greater than $m_ec^2$) to survive.

\vskip 0.3 true cm
\noindent
{\bf Relativistic motion ---}
If the source is moving relativistically, then the observed photon energies
are blueshifted, and the typical angles (as observed in the lab frame)
between photons are smaller, decreasing the probability for them to interact.
This solves the compactness problem.
Bulk Lorentz factors $\Gamma>100$ are required to avoid strong 
suppression of high energy $\gamma$--rays due to photon--photon collisions.
It is worth stressing here that the photon--photon problem
is particularly demanding if one associates the $>100$ MeV photons
seen by EGRET with the prompt emission of GRBs.
It is much less severe if it is instead associated with the
early afterglow. This is still an unsettled issue.

There is however a second argument demanding for strong relativistic motion,
concerning the very fast observed variability.
In fact the size associated with one millisecond is $R\sim 3\times 10^7$cm,
which is much too small to be optically thin.
To match the observed timescales with the size at which 
the fireball becomes transparent ($R_t\sim 10^{13}$ cm)
we need a Doppler contraction of time given approximately by 
$ct_{var} \sim R_t (1-\beta)$, yielding $\Gamma \sim 400$.


\subsection{The internal/external shock scenario}

If the central engine does not produce a single pulse, but works 
intermittently, it can produce many shells (i.e. many fireballs)
with slightly different Lorentz factors.
Late but faster shells can catch up early slower ones,
producing shocks which give rise to the observed burst emission.
In the meantime, all shells interact with the interstellar medium, and 
at some point the amount of swept up matter is large enough to decelerate
the fireball and produce other radiation which can be identified with 
the afterglow emission observed at all frequencies.

This is currently the most accepted picture for the burst 
and afterglow emission,
and it is called the internal/external shock scenario (Rees \& M\'esz\'aros 
1992; Rees \& M\'esz\'aros 1994; Sari \& Piran 1997).
According to this scenario, the burst emission is due to
collisions of pairs of relativistic shells (internal shocks), while
the afterglow is generated by the collisionless shocks produced by shells
interacting with the interstellar medium (external shocks).
All the radiation we see is believed to come from the 
transformation of ordered kinetic energy of the fireball into random energy.
For internal shocks, this must happen at some distance
from the explosion site, to allow the shells to be transparent 
to the produced radiation ($R_t \sim 10^{13}$ cm).
For external shocks, the deceleration radius, where the fireball starts
to emit the afterglow, depends on the density of the interstellar
medium (and by the possible presence of a stellar wind),
by the energy of the fireball and its bulk Lorentz factor.
For densities of the order of 1--10 proton/cm$^{3}$,
we get $R\sim 10^{16}$ cm as a typical value for the 
start of the afterglow.
Note that this would correspond to observe the initial
afterglow $\sim 150 R_{16}/\Gamma_2^2$ seconds after the trigger.

\subsection{Efficiency}

In the internal shock scenario we have collisions
of pairs of shells which are both relativistic,
with bulk Lorentz factors $\Gamma_1$ and $\Gamma_2$.
After the collision, the merged shell is still relativistic, with 
a bulk Lorentz factor which is between $\Gamma_1$ and $\Gamma_2$. 
The energy which is liberated in the process is therefore
a small fraction of the initial one, unless the ratio
$\Gamma_1/\Gamma_2$ is huge (Beloborodov 2000, Kobayashi \& Sari 2001).
But if the Lorentz factor of the shells is distributed
in a large interval, then a very fast shell would move in the
photon field created by the previous collisions, would scatter 
these ambient photons and would produce very high energy 
$\gamma$--rays by the inverse Compton process.
This {\it Compton drag effect} can be relevant: it can in fact 
decelerate the fast shells and then it would narrow the range of the 
bulk Lorentz factors of the shells undergoing internal shocks, 
thus lowering their efficiency (Lazzati, Ghisellini \& Celotti 1999).

\subsection{Radiation mechanisms}

It is reasonable to assume that internal and external shocks can 
amplify seed magnetic fields and accelerate electrons to relativistic 
random energies.
These are the basic ingredients for the synchrotron process,
which is therefore a strong candidate for the origin of the 
observed radiation of both the prompt and the afterglow emission.
%
There is indeed strong evidence that this is the main process
operating during the afterglow: the power law decay of the flux
with time, the observed power law energy spectra 
and the recently detected linear optical 
polarization in GRB 990510 (Covino et al. 1999; Wijers et al. 1999) 
and GRB 990712 (Rol et al. 2000).
The synchrotron nature of the prompt emission is instead controversial,
and alternatives have been proposed, such as quasi--thermal Comptonization
(Liang 1997, Ghisellini \& Celotti 1999a), Compton drag 
(Lazzati et al. 2000), relativistic inverse Compton emission
(Panaitescu \& Meszaros 2000) and ``jitter" radiation 
(i.e. electrons emitting by following magnetic field lines 
highly tangled on small scales; Medvedev 2000).
%

\section{Spheres versus jets}

A hot issue in the GRB field is the possible collimation of their
emitting plasma, leading to anisotropic emission able to 
relax the power requirements, at the expense of an increased
burst event rate.
In this respect polarization studies could be crucial, since 
there can be a link between the deceleration of a {\it collimated} 
fireball, the time behavior of the polarized flux and its position 
angle, and the light curve of the total flux.

\begin{figure}[h]
\vskip -0.5 true cm 
\centerline{\psfig{figure=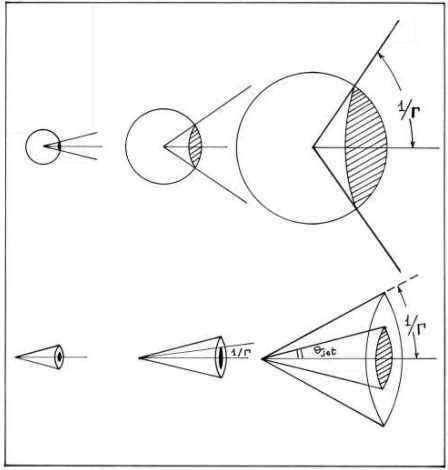,width=6cm}} 
\caption{Spheres or jets (or, rather, flying pancakes)?
This figure tries to explain a possible way to discriminate
between the two possibilities.
During the initial phases of the afterglow, the bulk Lorentz
factor is large, and consequently the observer sees only the 
fraction of the emitting area inside a cone with aperture
angle $\sim 1/\Gamma$.
There is no difference between a sphere and a jet during this phase.
In the spherical case the emitting area continues to increase
both because the radius of the sphere increases and because
$\Gamma$ decreases, allowing more surface to be within 
the $1/\Gamma$ cone.
In the case of collimation in a jet, once $1/\Gamma$ becomes
comparable to the jet opening angle $\theta$, the
observed surface increases only because the distance
to the jet apex increases. 
The light curve predicted in the two cases is therefore the same 
at early times, but in the jet case there will be a break at a 
particular time (when $1/\Gamma\sim \theta$), after which the 
light curve decreases more rapidly than in the spherical case.
}
\end{figure}
\subsection{Arguments in favor of collimation}
\begin{figure}[h]
\vskip -1. true cm 
\centerline{\psfig{figure=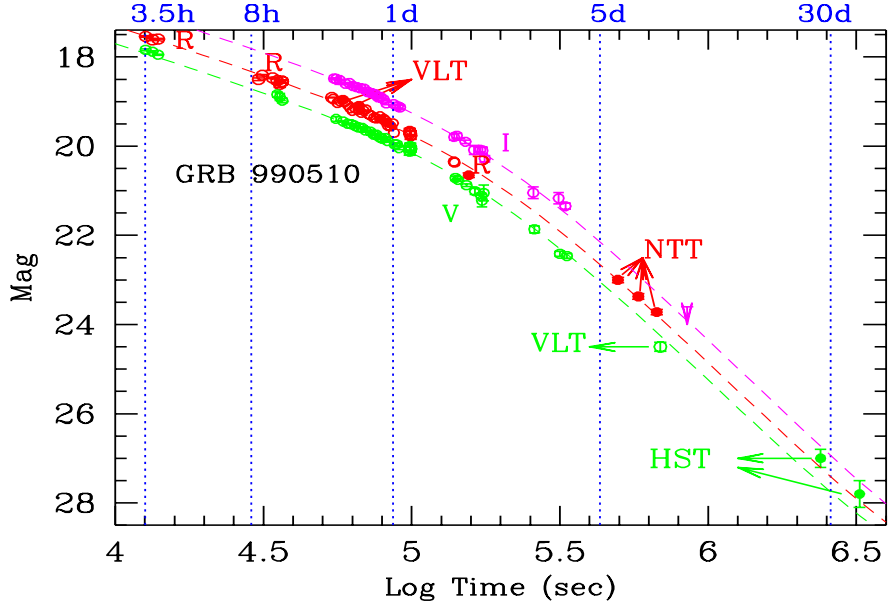, width=14cm, height=12cm}}
\vskip -3.5 true cm 
\caption{Light curve in different optical bands of GRB 990510,
the best example of jetted fireball. 
Note that the same curve fits the light curve
at different frequencies: the (smooth) break appears 
achromatic.  From Israel et al. (1999).}
\end{figure}
\begin{figure}
\vskip -1 true cm
\centerline{\psfig{figure=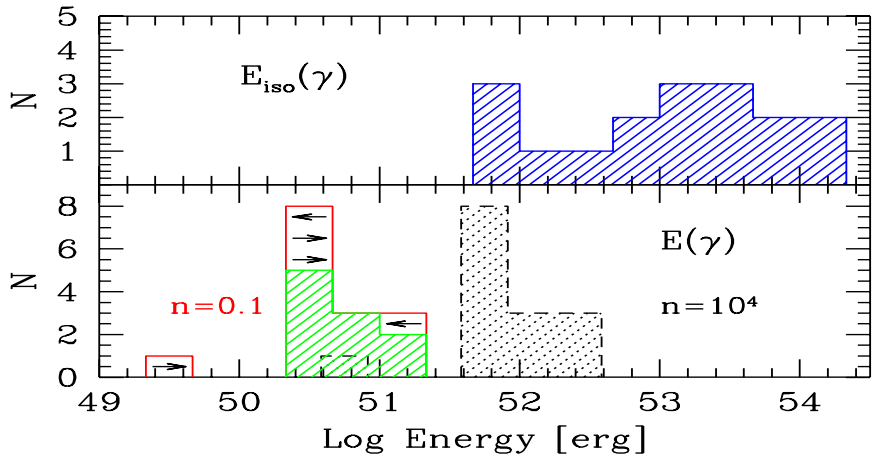, width=14cm, height=10.5cm}}
\vskip -4 true cm
\caption{The energy received in $\gamma$--rays from GRBs of known
redshift in the case of isotropic emission (upper panel),
and the ``true" energy if the emission is collimated in a cone, whose aperture
has been estimated by Frail et al. 2001
from the break in the light curve of the afterglows (bottom
panel), assuming a ISM density of $n=0.1$ cm$^{-3}$.
See how the ``true" energies are, on average, a factor $\sim 500$
lower than the ``isotropic" values. 
Also shown is the distribution of energies if the ISM density is $n=10^4$ 
cm$^{-3}$. Adapted from Frail et al. (2001).}
\end{figure}

\vskip 0.3 true cm
\noindent
{\bf Breaks in the light curves ---} 
Assume that the burst is collimated within a cone of semiaperture
$\theta$.
Assume also that, initially, the bulk Lorentz factor of the fireball is
such that $1/\Gamma <\theta$.
In this case, because of relativistic aberration, 
the observer (which is within the cone defined by $\theta$) 
will receive light only from a section of the emitting surface, 
of aperture $1/\Gamma$ and radius $R/\Gamma$, where $R$ is the 
distance from the apex of the cone.
This is illustrated in Fig. 12.
Initially, this area increases {\it both} because $R$ increases with
time {\it and} because $\Gamma$ decreases.
This leads to the estimate of how the received flux varies in time.
If the fireball is spherical, this will continue as long as the
motion is relativistic.
But if the fireball is collimated, there is a time when $1/\Gamma$
becomes comparable to $\theta$.
After this time the observed area will increase only because $R$ increases
while the decrease in $\Gamma$ will not ``enlarge" the available surface
(we ignore here the complications due to side expansion, which does not
change qualitatively the argument).
Since the rate of increase of the observed emitting area changes,
then there will be a change in the slope of the light curve.
An {\it achromatic} break is predicted.
The best example observed so far of such a case is the afterglow 
of GRB 990510, shown in Fig. 13.

\vskip 0.3 true cm
\noindent
{\bf Clustering of fireball energies ---} 
Taking advantage of the relation between the time of the break in the 
light curve and the degree of collimation (strong collimation should produce
an earlier break than mild collimation),
Frail et al. (2001) calculated the ``jet angle"
for a small sample of bursts of known redshifts.
This immediately yields the ``true" energy dissipated during the
prompt emission.
The remarkable results of Frail et al. (2001) is that 
despite the ``isotropic" energy values differ by some orders
of magnitude, the corrected values are all very similar 
and cluster around a value of a few times $10^{50}$ erg,
as shown in Fig. 14.
These values are however obtained assuming a density
of the decelerating interstellar medium of $n=0.1$ cm$^{-3}$,
a particularly small value, but in agreement with fits
to the afterglow spectral energy distribution (from radio to X--rays,
see, e.g. Panaitescu \& Kumar 2001).
Assuming a larger density makes the jet angle and the ``true" energy
to increase (even if they remain clustered: Fig. 14 shows how 
the distribution of energy shifts assuming $n=10^4$ cm$^{-3}$).

\vskip 0.3 true cm
\noindent
{\bf Polarized afterglows ---}
To produce polarized light some {\it asymmetry} is required.
If the radiation is due to the synchrotron process,
the magnetic field cannot be completely tangled, but must have some
degree of order within the emitting available (i.e. within the $1/\Gamma$
angle) volume. 
Even in a spherical source there can be 
distinct regions of ordered magnetic fields producing a net
polarized flux (Gruzinov \& Waxman 1999),
but a perhaps more natural asymmetry corresponds to a jet observed off axis
(note that the probability of observing a jet exactly on axis is 
vanishingly small).
%
Ghisellini \& Lazzati (1999) and Sari (1999) have  
considered a jet geometry where the initially tangled magnetic field
is  ``squeezed" (by compression) in one direction.
It appears completely random for face--on observes 
(with respect to the direction of compression), but highly ordered to 
edge--on observers (a similar model was proposed by Laing 1980
for polarized, radio--loud Active Galactic Nuclei). 
Photons emitted in the plane of the slab can then be highly polarized.
If the slab moves with Lorentz factor $\Gamma$,
those photons emitted in the slab plane (perpendicularly to the
direction of motion) are aberrated in the observer frame, and make an angle
$\theta =1/\Gamma$ with respect to the slab velocity.
Observers looking at the moving slab at this angle will detect a large degree 
of optical polarization.
In particular there will be 2 maxima in the light curve of the 
polarized light (which become 3 considering side expansion),
with a switch of 90 degrees in the polarization angle between the two 
maxima.

Should these predictions be confirmed, we would have a very powerful tool to 
know the degree of collimation of the fireball, and hence the true total 
emitted power.

%

\begin{figure}
\centerline{\psfig{figure=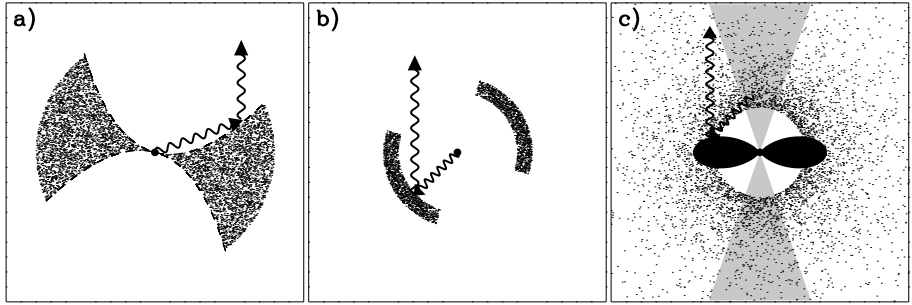, width=12cm}} 
\caption{Scenarios for some of the models proposed to explain
the broad iron line emission of GRB 991216. 
From left to right:
a): in this scenario there is a sort of plerion  with an excavated funnel
illuminated by a photoionizing X--ray source. The walls of the
funnel reprocess the ionizing flux and produce the iron line.
b): an asymmetric young supernova remnant illuminated by some
ionizing flux. In this case the observer sees radiation and the 
iron line coming from the interior of the remnant.
c): the burst is surrounded by some scattering material (e.g. the
pre--hypernova wind) which reflects back some of the burst and 
afterglow photons. 
These photons are intercepted by the material in the envelope
of the star, which is expanding sideways,
and reprocessed to form the iron (or Cobalt, or Nickel) line. From 
Vietri et al. (2001). }
\end{figure}

\subsection{Arguments against strong collimation: Iron lines }

The presence of (possibly broad) iron lines in the X--ray afterglow 
of some bursts imply that a dense, possibly iron--rich
material must be present in the vicinity of the burst site.
The large densities involved suggest that the line emission process
is fast photoionization and recombination by an optically thick slab,
reprocessing the impinging ionizing continuum in its $\tau\sim 1$ layer
(where $\tau$ is the relevant optical depth).
The observed line flux would then be proportional to the projected
emitting area, which becomes very small if the burst and afterglow fluxes
are collimated in a cone as narrow as envisaged, e.g. by Frail et al. (2001).

\vskip 0.3 true cm
Models proposed so far for the iron line origin can be broadly
divided into two categories: ``internal" (Meszaros \& Rees 2001) 
and ``external" (Lazzati, Campana \& Ghisellini 1999; 
Vietri et al. 2001) models.
The latter ones assume that a supernova explodes some time before the 
burst and forms a dense and iron rich remnant at $10^{15}$--$10^{16}$
cm from the burst. 
This material is illuminated by the burst and afterglow X--ray flux,
which can also transfer some linear momentum to the walls of the 
funnels excavated inside the remnant (see Fig. 15) which is therefore
accelerated to velocities of $\sim 10^4$ km s$^{-1}$.
This can help explaining a paradox:
in GRB 991216 the line is broad, implying velocities of 
15,000 km s$^{-1}$ (Piro et al. 2000). 
Since we can estimate (Eq. 1) the size of the emitting material,
we can also estimate the time elapsed from the supernova explosion,
and this turns out to be much too short for the decay of
Cobalt and Nickel into iron (Vietri et al. 2001).  

In the ``internal" models the reprocessing material is identified
with the funnel walls or the envelopes of the hypernova thought
to be the progenitor of the burst.
In such a scenario the super/hyper-nova explodes
at the same time of the burst.
In this case the appropriate density can be obviously much greater then
in the ``external" model, enhancing the recombination rate
and therefore requiring less iron.
On the other hand one needs a continuous illuminator lasting for at least
one day after the burst explosion, generating $\sim 10^{47}$ erg s$^{-1}$
to sufficiently photoionize the iron atoms.
Three possibilities have been proposed: 1) instead of a black hole,
the compact object resulting from the burst explosion is a magnetar,
with a fastly decaying magnetic field;
2) accretion by fall--back material onto the black hole;
3) side--deposition at the top of the funnel of $\sim 10^{51}$ erg
producing a continuous X--ray radiation illuminating filaments or
clumpy material which generates the line.

We can see that in both class of models there must be a sizeable
fraction of energy not well collimated (e.g. the relevant $\theta$
must be relatively large).

There is finally a very serious energy budget problem: 
if the ``true" energy derived by Frail et al. (2001) is to be taken seriously,
we have a few times $10^{50}$ erg
available in $\gamma$--rays, and a factor $\sim 20$ less
(i.e. $E_{ioniz}\sim 10^{49}$ erg) in ionizing X--ray photons.
The observed lines have $L_{Fe}\sim 10^{44}$ erg s$^{-1}$.
If they last for $10^5$ seconds we have $E_{Fe}\sim E_{ioniz}$:
that is, a completely unreasonable conversion efficiency of 100 per cent.

\section{The primary energy source}

The greatest unknown in GRB science is what is the 
progenitor.
In recent years the attention of the community polarized onto three 
proposals: 
1) the merging of two compact objects, such as
two neutron stars forming a black hole surrounded by some
accreting neutron--dense torus (Paczynski 1986; Goodman 1986;
Eichler et al. 1989; Meszaros \& Rees 1997);
2) the core collapse of a very massive star ({\it hypernova},
Woosley 1993; Paczynski 1998; Fryer, Woosley \& Hartmann 1999);
3) the formation of a black hole from a rapidly spinning but decelerating
neutron star left over by a previous explosion of a ``quasi--normal"
supernova ({\it supranova}, Vietri \& Stella 1998).

Irrespective of the differences among these three proposals, 
the central engine could be very similar,
being composed by the same ingredients: a fast spinning black hole
surrounded by a very dense (neutron--dense) torus.
In this case the available sources of energies, apart from gravitational
radiation, are in the form of neutrinos, accretion of the torus material
onto the black hole, and rotation of both the torus and the hole.

\begin{figure}
\vskip -1 true cm 
\centerline{\psfig{figure=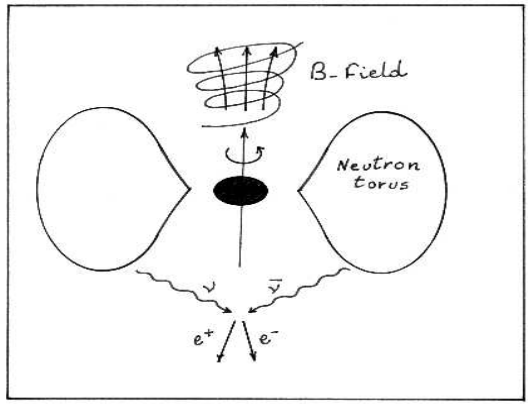,width=8cm, height=6cm}}
\caption{The basic GRB engine.
Irrespective of the progenitors, the final scenario for the
extraction of energy powering the GRB could be the one sketched
in this figure.
A neutron--dense torus is orbiting around and accreting onto a 
(possibly newly formed) spinning black hole.
During the formation of both the hole and the torus, 
a fraction of a solar mass--energy is converted into neutrinos,
which could interact between themselves and form electron--positron pairs.
Most of the energy is however in the form of rotational energy
of the black hole.
The Blandford--Znajek mechanism could extract it, provided that
a sufficiently strong ($\sim 10^{15}$ Gauss) magnetic field  
threads the hole.
}
\end{figure}

The black hole is more massive than the disk: it should have at least
2 solar masses, and probably more in the collapsar scenario,
versus 0.1--0.2 solar masses in the torus.
Therefore, if the hole is fastly spinning, most of the energy
is in its rotation.
One promising way to extract this energy  is the Blandford \& Znajek (1977) 
process, in which the rotational energy of a Kerr black hole 
can be extracted by a magnetic field surrounding the hole providing
a source of power:
\begin{equation}
L_{\rm BZ}\,  \sim \, 10^{51}\, \left({a\over m}\right)^2\,
\left({M_{\rm BH} \over 10\, M_\odot}\right)^2
\left({B\over 10^{14}{\rm G}}\right)^2\,\, {\rm erg\, s^{-1}} \,  
\sim \, \left({ a\over m}\right)^2  (3R_s)^2 U_B c
\end{equation}
where $(a/m)$ is the specific black hole angular momentum 
($\sim 1$ for maximally rotating black holes), $R_s$ is the Schwarzchild 
radius and $U_B=B^2/(8\pi)$.
For a maximally rotating black hole of 10 solar masses the extractable 
rotational energy is 
$0.29\times 10 M_\odot c^2\sim 5\times 10^{54}$ erg:
even if the bursts are not collimated in a narrow cone, 
there is (theoretically) plenty of energy stored in the black hole spin.

\section{Open issues and problems}
In the very recent past the GRB field witnessed extraordinary
successes both in the theory and in the observations,
and the fireball--internal/external shock scenario
is becoming a paradigm.
All this is obviously great, since a framework is always
useful to systemize ideas, observations, and maybe to sharpen 
our criticisms.
On the other hand a paradigm helps lazy people to do useful 
but ``ordinary" work without exploring alternatives.
Therefore it can be useful to point out some areas where the proposed 
solutions are still not completely convincing (even if this will be a 
biased list, based on my own prejudices).

\vskip 0.3 true cm
\noindent
{\bf Efficiency and energetics of internal shocks ---} 
Kumar (1999); Lazzati, Ghisellini \& Celotti (1999);
Spada, Panaitescu \& Meszaros (2000) and Panaitescu \& Kumar (2001)
pointed out that internal shocks are not the most 
efficient way to transform the bulk kinetic energy of the fireball(s)
into radiation.
With a Lorentz factor contrast of a few between the colliding shells
we have efficiencies of the order of 1 per cent.
This implies that: 1) the total fireball energy is much more
than what we estimate from the $\gamma$--ray radiation, and
2) that the afterglow, powered by the much more efficient external 
shocks, should radiate more energy than the burst.
This is not not what we observe.

\vskip 0.3 true cm
\noindent
{\bf The peak energy of the prompt emission ---} 
The peaks (in $EF_E$) of the prompt emission spectra ($E_{peak}$) are 
remarkably narrowly distributed, clustering around a few hundreds keV.
This despite the possible spread of the Lorentz factor values
and also the different redshifts.
The value of $E_{peak}$ is tantalizingly close to $m_ec^2$, possibly
suggesting the importance of Compton down--scattering and/or 
photon--photon processes for the prompt emission
(Ghisellini \& Celotti 1999b; Thompson \& Madau 2000; Brainerd 1994).

\vskip 0.3 true cm
\noindent
{\bf Radiation mechanism of the prompt emission ---}
In the internal shock scenario, the prompt emission is thought to 
be synchrotron emission by electrons with a low energy cutoff.
This should yield spectra with a low energy tail $F(E)\propto E^{1/3}$
(equivalent to $N(E)\propto E^{-2/3}$),
in rough agreement with the distribution of the low energy 
spectral index $\alpha$ (as shown in Fig. 6).
But in the very same scenario, the magnetic field is large enough
to make electrons radiatively cool on a timescale much shorter
than a milliseconds (and of course much shorter than any
integration time to get a spectrum).
Furthermore, we {\it know} that the electrons {\it must} cool fast,
since the light curves do vary on the millisecond timescales.
The synchrotron spectrum of a cooling electron is $F(E)\propto E^{-1/2}$
(equivalent to $N(E)\propto E^{-3/2}$)
and this slope is much too softer than observed 
(Ghisellini, Celotti \& Lazzati 2000).

\vskip 0.3 true cm
\noindent
{\bf What causes the intermittent release of energy ---}
The term ``fireball" may lead to think of a great and single explosion.
But what we need in GRBs is an intermittent source of energy
(i.e. thousands of fireballs).
Consider that the Schwarzchild radius light crossing
timescale for a 10 solar mass black hole is $10^{-4}$ s, 
while we see bursts lasting for more than 100 seconds.
To make a comparison, think of a $10^9 M_\odot$ AGN black
hole producing a very energetic phenomenon lasting for  300 years.
Indeed, there is time for quasi--stationary processes.
Even if the Blandford--Znajek process can indeed work
for ``all" this time (and for hundreds million years in radio--loud AGNs),
it is not clear what can cause the on--off states as witnessed
by the erratic light curves.

\vskip 0.3 true cm
\noindent
{\bf What fixes the relatively small amount of mass in the fireball ---}
We are convinced that the bulk Lorentz factors are of the order
of $\Gamma\sim 100$.
This must correspond to a moving mass of the order of $10^{-6}$--$10^{-4}$
solar masses (depending on the ``true" fireball energy).
For a fixed value of the fireball energy per unit solid angle,
the above amount of mass cannot be much less, since otherwise the fireball 
becomes transparent when pairs annihilate, letting internal
radiation escape freely with a blackbody spectrum and leaving
no energy for the afterglow.
It cannot be much more than that, since otherwise the bulk Lorentz
factor becomes too small to account for the observed variability.
Therefore we must explain what causes the ``right" amount
of barion contamination.


\vskip 0.3 true cm
\noindent
{\bf Density of the circumburst material ---}
The three leading models mentioned in Section 7 envisage very different
densities of the circumburst material.
In fact in the collapsar model the GRB is associated with a
newly formed star, which should live in a star forming region
of densities of about $10^{4\pm 1}$ cm$^{-3}$, corresponding, for
a 1 pc region, to Hydrogen column densities of $N_H\sim 3\times 10^{22\pm 1}$ 
cm$^{-2}$.
Besides that, the star itself should be surrounded by the 
circumburst material of its stellar wind. 
Studies of winds in massive stars give outflows of
$10^{-5}$--$10^{-4}$ $M_\odot$ yr$^{-1}$, with a velocity of 
$10^2$--$10^3$ km s$^{-1}$.
Supranovae should be located in dense star forming region as well, 
but during the delay between the supernova and the burst explosion,
the newly born neutron star can ``clean" the environment through the
super--Eddington luminosity produced by the neutron star before
the GRB event (Vietri \& Stella 1998).
Finally, NS--NS mergings should instead happen far away from 
their original birth place, and so be characterized by 
surrounding densities of the order of 0.1--1 cm$^{-3}$.

Therefore it seems odd that there is some consensus regarding 
collapsars as the progenitor for long bursts, and at the same time
almost all hints we currently have indicate a small value for the density,
i.e. 0.1--10 cm$^{-3}$.
These include fits of the afterglow spectra 
(see e.g. Panaitescu \& Kumar 2001) 
and X--ray absorption rarely exceeding $N_H=10^{21}$ cm$^{-2}$ 
(Owens et al. 1998; Galama \& Wijers 2001; Stratta et al., in prep.).
This issue could greatly benefit from early low energy X--ray 
observations, revealing a time dependent $N_H$ column, since the absorbing
material, being neutral at first, can become completely ionized 
after the passage of the burst photons if sufficiently
close to the GRB site (Lazzati \& Perna 2001).

\begin{figure}[h]
\vskip -1 true cm 
\centerline{\psfig{figure=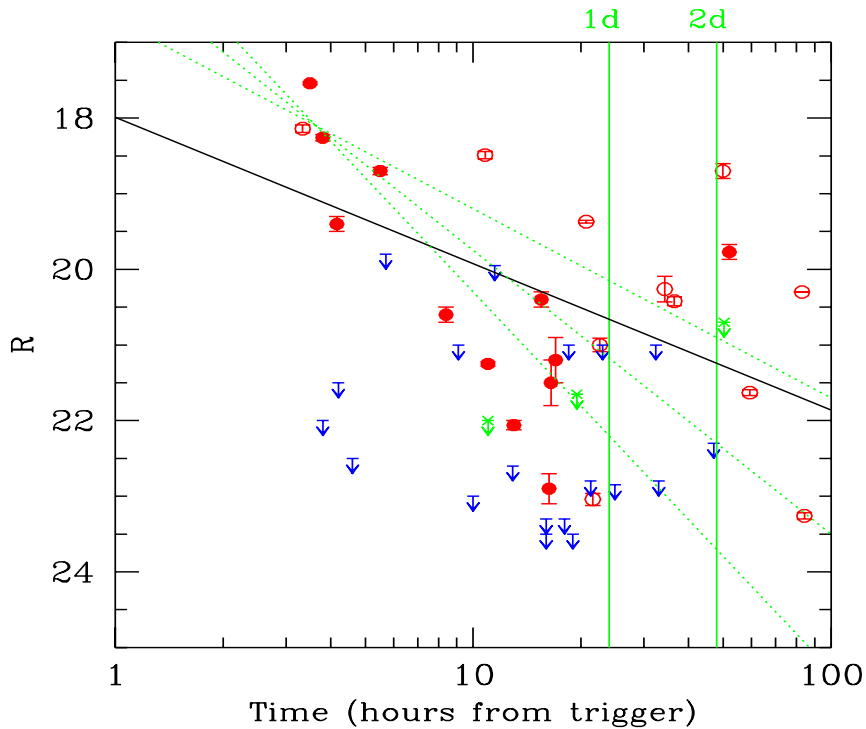,width=13cm, height=11cm}}
\vskip -1 true cm 
\caption{Detection $R$ magnitude (or upper limits) versus the time 
of observation for a set of afterglows. 
The comparison between upper limits (arrows) and detection
(circles) shows that the non--detection of the optical afterglows
is not due to too short exposure time or too late observations. From 
Lazzati, Covino \& Ghisellini (2002).
}
\end{figure}

\vskip 0.3 true cm
\noindent
{\bf ``Failed optical afterglows" ---}
More than half of the bursts with detected X--ray afterglows
were not detected in the optical.
This is {\it not} due to adverse observing conditions
(i.e. too short exposures or too late observations),
as shown by Fig. 17 (see Lazzati, Covino \& Ghisellini 2002).
Are they optically reddened? 
If the dust responsible for absorption is too close to the GRB site 
(within a few parsec) it should sublimate (Fruchter, Krolik \& Rhoads 2001),
while if it is far away (as in a ultra--luminous IR galaxy, Ramirez--Ruiz, 
Trentham \& Blain 2001) we should see some absorption in X--rays.
The other alternatives are very high redshift bursts ($z>5$, so that 
the Ly--$\alpha$ clouds absorb all of the optical flux),
or an intrinsic optical weakness, despite of a normal behavior
in X--rays.

\section{The future}
It is my impression that the field of GRB is now living the same
excitement felt soon after the discovery that quasars were at
cosmological distances.
If true, then the GRB field should be in its infancy, even though
the next four or five years could bring it to adultness.
This is because we expect a huge impetus from new planned
mission and projects, and because the field has already
attracted hundreds of researches.

\subsection{Issues}

\vskip 0.3 true cm
\noindent
{\bf Short bursts ---}
Are short bursts really a separate class of bursts,
with a different progenitor?
We need accurate localization of short bursts
and of course to find their redshifts.
{\it Beppo}SAX only triggered on long bursts, so
our hope relies on HETE II and/or the next missions.

\vskip 0.3 true cm
\noindent
{\bf Cosmology ---}
If bursts are really associated with massive stars, then
they should trace the high mass end of the initial mass function
at all redshifts.
There is the exciting possibility for a small but
important fraction of bursts to be associated to the 
so called Pop III stars, thought to be responsible of the 
first metal enrichment of the primordial gas
and possibly for its re--ionization 
(see e.g. the review by Loeb \& Barkana 2001).

\subsection{Projects}

\vskip 0.3 true cm
\noindent
{\bf Swift ---}
The most powerful tool we will have to study GRB is the dedicated
satellite called Swift ({\tt http://swift.gsfc.nasa.gov/}).
It will host three instruments: a hard X--ray [10--150 keV] coded mask
with a wide field of view and localization accuracy of a few arcminutes,
a low and medium X--ray telescope [0.1--10 keV] with arcsecond
localization performances, and an optical monitor (30 cm of diameter)
with optical and UV filters and low dispersion grisms.
Swift is planned to observe 250--300 burst per year,
to distribute their coordinates immediately, to slew to target 
in a few tens of seconds, and start to observe with the X--ray mirrors
and the optical monitor while the prompt emission can still be on.
The launch is foreseen in 2003.
All data will be public.

\vskip 0.3 true cm
\noindent
{\bf REM: an infrared robotic telescopes ---}
On ground, even before Swift, several groups are planning to
organize the necessary follow--up observations of GRBs,
to find their redshift and also to observe at other wavelengths,
not covered by the optical monitor onboard Swift.
This will in fact be blind in the infrared, where dust 
and/or Ly--$\alpha$ clouds absorption have small or no effect.
It is therefore crucial to have a completely robotic,
fast slewing infrared telescope, such as
REM (Rapid Eye Mount, see the web page at:\\
 {\tt http://golem.merate.mi.astro.it/projects/rem/}).\\
This will also be equipped with a dichroic beam splitter, and it will
be able to perform simultaneous IR and optical spectrophotometric
observations.
If the afterglow of a burst is seen in the IR but not in the optical,
we will have a very good high redshift candidate object,
and we will have the opportunity to alert much larger telescopes
(REM has a 60 cm diameter mirror) to perform high resolution
infrared spectroscopy at potentially any redshift.
{\it There is no other way to acheive this result}.


\section*{Acknowledgments}
I thank Annalisa Celotti and Davide Lazzati for years
of fruitful collaboration and them again, together
with Sergio Campana, Stefano Covino, and Daniele Malesani
for their help in improving this manuscript.


\end{document}